\documentclass[12pt]{article}
		\title{Use of Historical Individual Patient Data in \\Analysis of Clinical Trials}

\pdfoutput=1
\date{}
	\usepackage{amsmath,latexsym, graphics,graphicx,showexpl,amsthm,amsfonts,fullpage}
\usepackage[left=1in,top=1in,right=1in,nohead]{geometry}
\usepackage{graphicx}
\usepackage[margin=10pt,font={small},labelfont=bf]{caption}
\usepackage{subcaption}
\usepackage{bbm} 
\usepackage{setspace}
\usepackage{natbib}
\usepackage[titletoc]{appendix}
\usepackage{epstopdf}
\usepackage[pdftex,bookmarks,colorlinks=true,linkcolor=blue]{hyperref}
\usepackage{algpseudocode,algorithm}
\usepackage{authblk}
\usepackage{multicol}
\usepackage{hyperref}
\usepackage{array}
\usepackage{etoolbox}
\usepackage{setspace}

\newcolumntype{L}[1]{>{\raggedright\let\newline\\\arraybackslash\hspace{0pt}}m{#1}}
\newcolumntype{C}[1]{>{\centering\let\newline\\\arraybackslash\hspace{0pt}}m{#1}}
\newcolumntype{R}[1]{>{\raggedleft\let\newline\\\arraybackslash\hspace{0pt}}m{#1}}
\AtBeginEnvironment{thebibliography}{\linespread{1.7}\selectfont}

\makeatother
\author[1]{Shirin Golchi\thanks{e-mail: shirin.golchi@mcgill.ca}}
\affil[1]{McGill University, Department of Epidemiology, Biostatistics and Occupational Health}
\begin{document}
	
{\linespread{1.5}	
	\maketitle

\begin{abstract}
Historical data from previous clinical trials, observational studies and health records may be utilized in analysis of clinical trials data to strengthen inference. Under the Bayesian framework incorporation of information obtained from any source other than the current data is facilitated through construction of an informative prior. The existing methodology for defining an informative prior based on historical data relies on measuring similarity to the current data at the study level and does not take advantage of individual patient data (IPD). This paper proposes a family of priors that utilize IPD to strengthen statistical inference. It is demonstrated that the proposed prior construction approach outperforms the existing methods where the historical data are partially exchangeable with the present data. The proposed method is applied to IPD from a set of trials in non-small cell lung cancer.
\end{abstract}

\noindent%
{\it Keywords: Historical controls; Informative prior; Power prior; Similarity measure.}  

}
\section{Introduction}
\label{sec:intro}
%The use of historical data to strengthen statistical inference appears in various settings and applications. A few examples are analysis of clinical trials data where information is available from observational studies, meta-analysis, and analysis of social survey results with information borrowed from other sources such as the census \citep{Gol17}. The focus of the present paper is informed analysis of clinical trials using external data sources. However, the proposed methodology can be well adapted to other settings.

Utilizing historical or external data to strengthen statistical inference in clinical trials has been explored in a variety of settings \citep{DemSelWee1981, Tar1982, IbrRyaChe1998}. Within the clinical trials framework, common scenarios include studies where achieving adequate power requires infeasibly large sample sizes. Sample size restriction can be either due to scarcity of the eligible population (rare diseases) or small effect sizes that are clinically meaningful but require larger than feasible study sizes to achieve statistical significance. Another interesting family of studies are trials where a control cohort is completely or partially absent for ethical or practical reasons. Single arm phase I or II trials have become popular in oncology and drug development; and the design and analysis of these trials require effective and robust statistical methodology in utilizing past studies \citep{ThaSim90, ZohTerZho08, CheCheMor16}.

Under the Bayesian framework, information from any  source other than the study data are incorporated via the prior distribution. A variety of approaches have been proposed in the literature for construction of informative priors based on historical data. One of the most popular methods is the power prior. Introduced formally by \cite{CheIbr2000}, power priors are based on a weighted log-likelihood of the historical data. The role of the weight (or power) is to control the level of contribution of past studies to inference. Various methods have been proposed for specification of the power(s). For a comprehensive review of theory and application of power priors see \cite{IbrCheGwo15}.

Another popular family of methods for taking advantage of historical and external data determine the amount of borrowed information through a parameter that estimates between study heterogeneity in a hierarchical modelling framework. Meta-analytic predictive priors (MAP) \citep{NeuCapBra2010, SchGstRoy14, RovFri19, LewSarZhu19, WebLiSea19} and commensurate priors \citep{HobSarCar12, MurHobLys2014} fall under this family of priors with different formulations of ``study effect".

Other work on prior construction based on the above mentioned methods include that of \cite{CheIbr06} who investigate the relationship between power priors and priors based on hierarchical modeling; \cite{HonFuCar18} who consider power priors and MAP for synthesizing aggregate and IPD in a network meta-analysis (NMA) framework; \cite{RosDejNor2018} who provide a review of a variety of information borrowing methods with comparisons in terms of design operating characteristics; and \cite{Gal2016} who discuss the bias variance trade-off for dynamic borrowing through hierarchical models.

While the above mentioned methods are powerful tools, they have a common limitation in using partial information (rather than partially using information) from historical studies. The amount of information incorporated into the prior depends on the aggregate level data rather than individual patient data. Under the power prior approach, for example, each study is given a power, based on its similarity measured on likelihood-based criteria, to the concurrent study. Information borrowing methods that rely on a hierarchical structure penalize heterogeneity in historical studies since the amount of information borrowed depends on the variance parameter that captures study (rather than individual patient) differences. As a result, less information is used from all of the studies even if some strongly resemble the concurrent data. 

%Consider, for example, the hypothetical scenario presented in Figure~\ref{fig:toy}. The concurrent study $S_c$ results are aimed to be generalized to the target population. Among the three available historical studies, the first one $S_1$ is assumed to be generalizable to the same population as the concurrent study and is, therefore, is expected to be fully exchangeable with the concurrent study in terms of the distribution of outcomes. The second study $S_2$ is a sample of a hypothetical population, B, that has a negligible overlap with the target population. This means that there may be individuals in $S_2$ whose characteristics match the target population but the majority of individuals do not belong to the target population. The hypothetical population, A, on the other hand, contains the target population -- for example is defined based on a wider age range of patients. Therefore, $S_3$ is expected to have a significant overlap with the concurrent study but there are individual patient data under $S_3$ that should not be included in the inference. 

Currently, no method of prior construction exists that allows the researcher to use a portion of individual patient data from a study according to individuals' propensity under the current study population. Under the existing approaches, a measure of similarity or target population membership is explicitly or implicitly estimated at the study level rather than the individual level. This results in loss of information in cases where the study populations partially overlap.

The goal of the present work is to address this gap in the literature by proposing a prior construction method that utilizes individual patient data (IPD). The proposed approach is a generalization of power priors such that every individual within the historical studies is assigned a distinct power (or weight). The powers are specified with respect to a distance or similarity measure to the target population. Assuming that the concurrent trial sample is the closest proxy for the target population, this distance measure is estimated as the propensity of individuals under the concurrent study population according to the joint distribution of available variables. A truncation threshold is proposed for the weights to prevent estimation bias due to large number of small contributions from unexchangeable potions of IPD.  

A simulation study is designed to assess the performance of the proposed priors in terms of estimation accuracy and precision as well as power in compare to select existing methods under different exchangeability scenarios between the historical and current studies. Exchangeability is defined as closeness of the joint distribution of all available variable - including covariates and outcomes - and can be violated by different covariate distributions or different outcome distributions that cannot be explained/adjusted for through the measured covariates. The proposed approach is applied to IPD from a set of clinical trials in non-small cell lung cancer (NSCLC).

The remainder of the paper is organized as follows: A motivating example is presented in Section~\ref{Sec:me}. The proposed methodology is described in Section~\ref{Sec:methods}. Section~\ref{Sec:sim} follows with a simulation study where the proposed approach is compared to existing prior construction methods. The proposed method is applied to the motivating example in Section \ref{Sec:app} and a discussion follows in Section \ref{Sec:dis}.

\section{Motivating Example}\label{Sec:me}
As a motivating example consider four trials in second line NSCLC with sufficiently similar patient population and the common primary outcome of overall survival: INTEREST \citep{INTEREST}, ZODIAC \citep{ZODIAC}, PROCLAIM \citep{PROCLAIM} and Study 57 \citep{Study57}. All trials were conducted for participants who had previously been treated for NSCLC.
The IPD data for these studies was acquired from Project Data Sphere (\href{http://www.projectdatasphere.com}{http://www.projectdatasphere.com}), an open-source repository of individual-level patient data from oncology trials. A brief summary of key trial characteristics for the included trials is provided in Table 1. 

\begin{figure}[t]

		\centering
		\includegraphics[width= \textwidth]{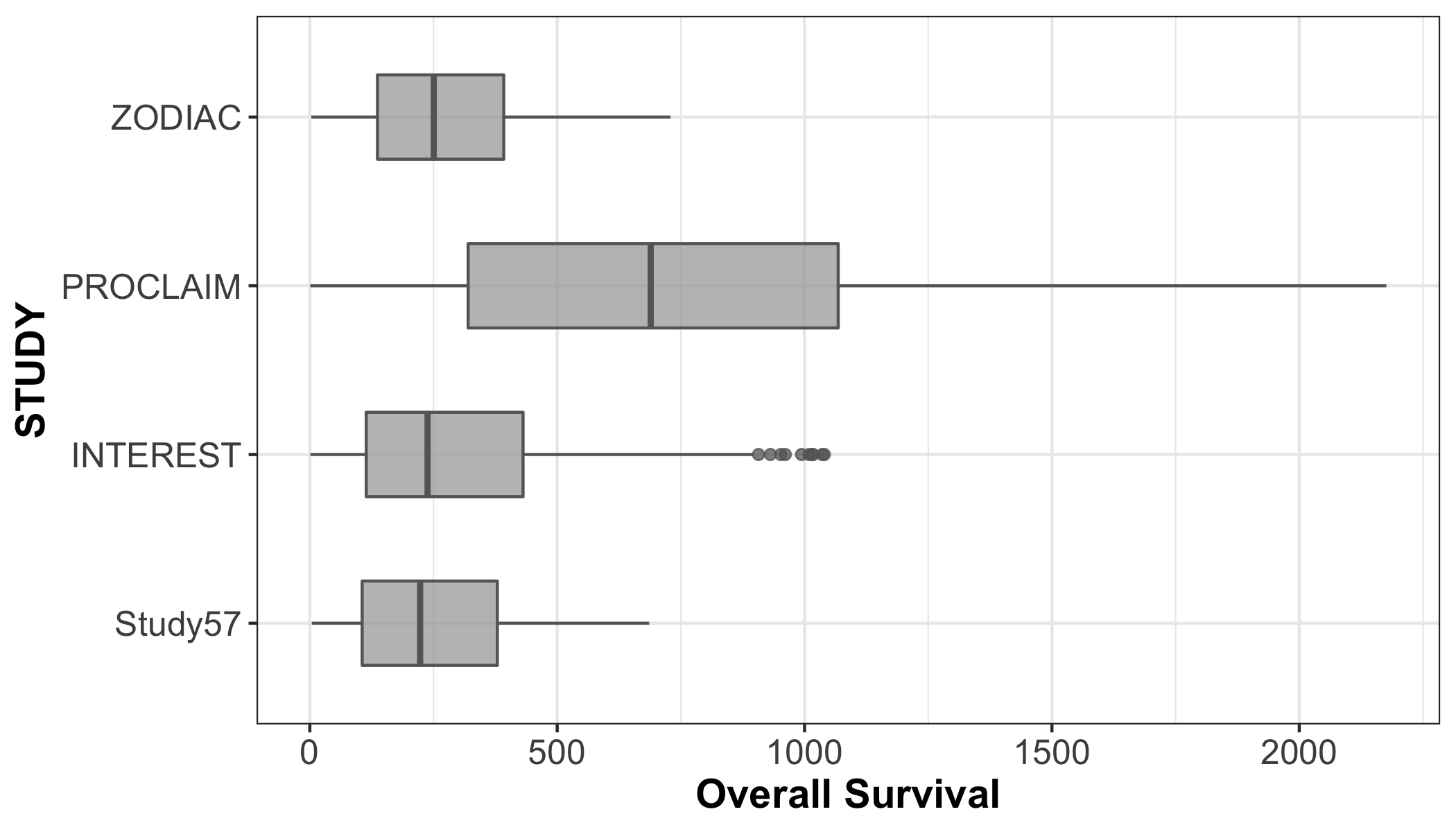}
	
	\caption{Overall survival distribution for the four NSCLC trials .}
	\label{fig:me}
\end{figure}

Patients within INTEREST, ZODIAC and Study 57 were predominantly stage IV, however PROCLAIM exclusively recruited  stage III patients (of which 52\% were stage IIIB). This is reflected in the control group median survival time which was between 8-10 months for all trials except for PROCLAIM that demonstrated a median survival time of 25 months (Figure~\ref{fig:me}). 

Other differences between studies included exposure to prior therapy. Whilst all included patients had previously received at least one previous chemotherapy regimen, the proportion of patients who had received two or more varied between 0 (PROCLAIM, ZODIAC) to 35\% (Study 57). Similarly, radiotherapy varied significantly, with PROCLAIM being the only trial which permitted (concurrent) chemoradiotherapy. Other patient characteristics were largely well balanced between groups, with an average age of $60 \pm 1$ years, and a majority of patients (54-75\%) having adenocarcinoma histology. 

Consider the hypothetical scenario that STUDY57 is the ``concurrent" trial and we are interested in using data from the other three clinical trials to enrich the control arm with the goal of improving inference and achieving more power. Given the brief description of the four trials provided above, PROCLAIM is substantially different that the three other trials and should not be used to inform the inference. However, we include this study in our analyses and comparisons for illustrative purposes.

\begin{figure}[t]
	
	\begin{subfigure}[b]{0.5\textwidth}
		\centering
		\includegraphics[width=\textwidth]{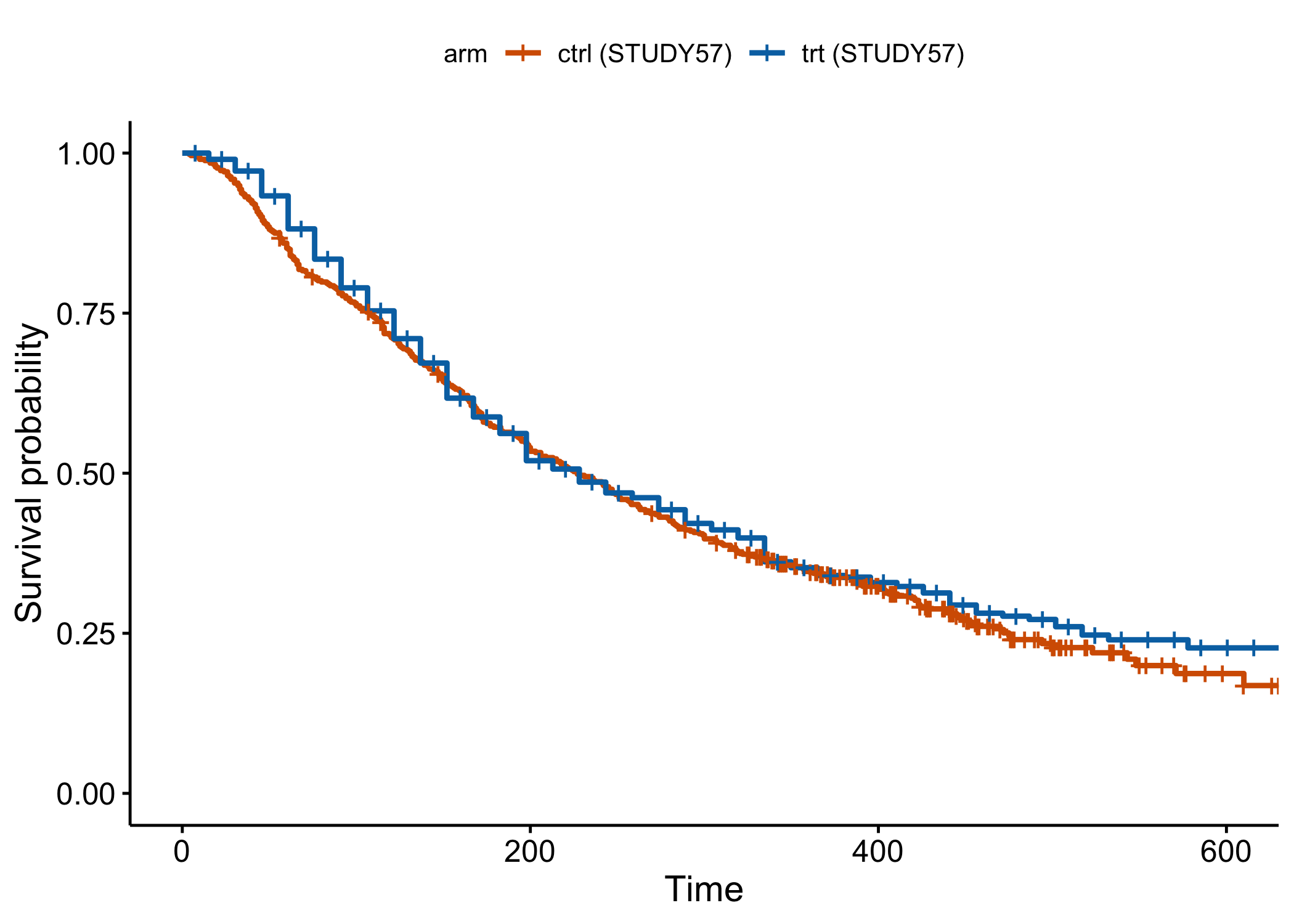}
		\caption{}
		\label{fig:KM11}
	\end{subfigure}
	\begin{subfigure}[b]{0.5\textwidth}
		\centering
		\includegraphics[width=\textwidth]{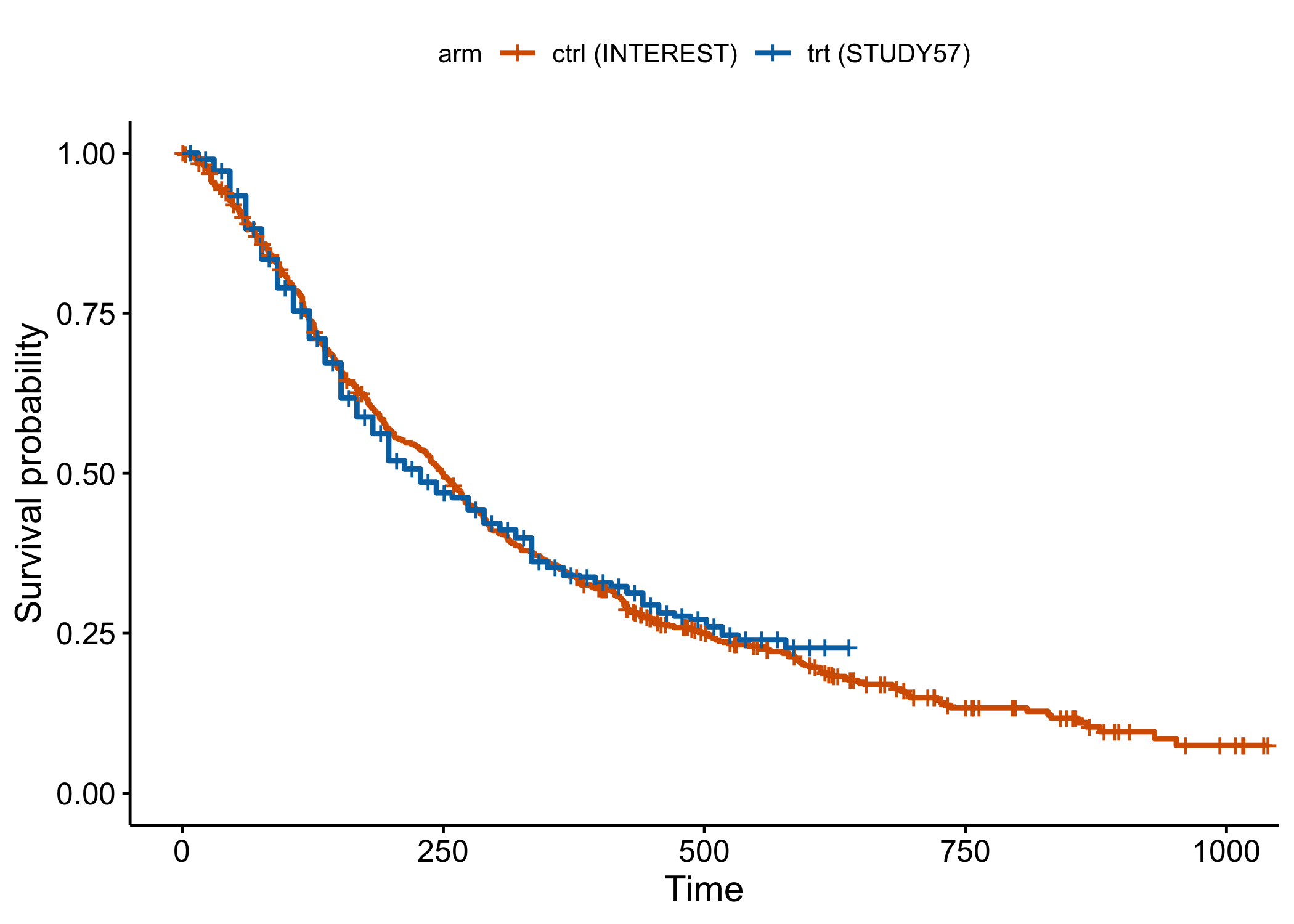}
		\caption{}
		\label{fig:KM12}
	\end{subfigure}
	
	\begin{subfigure}[b]{0.5\textwidth}
		
		\includegraphics[width=\textwidth]{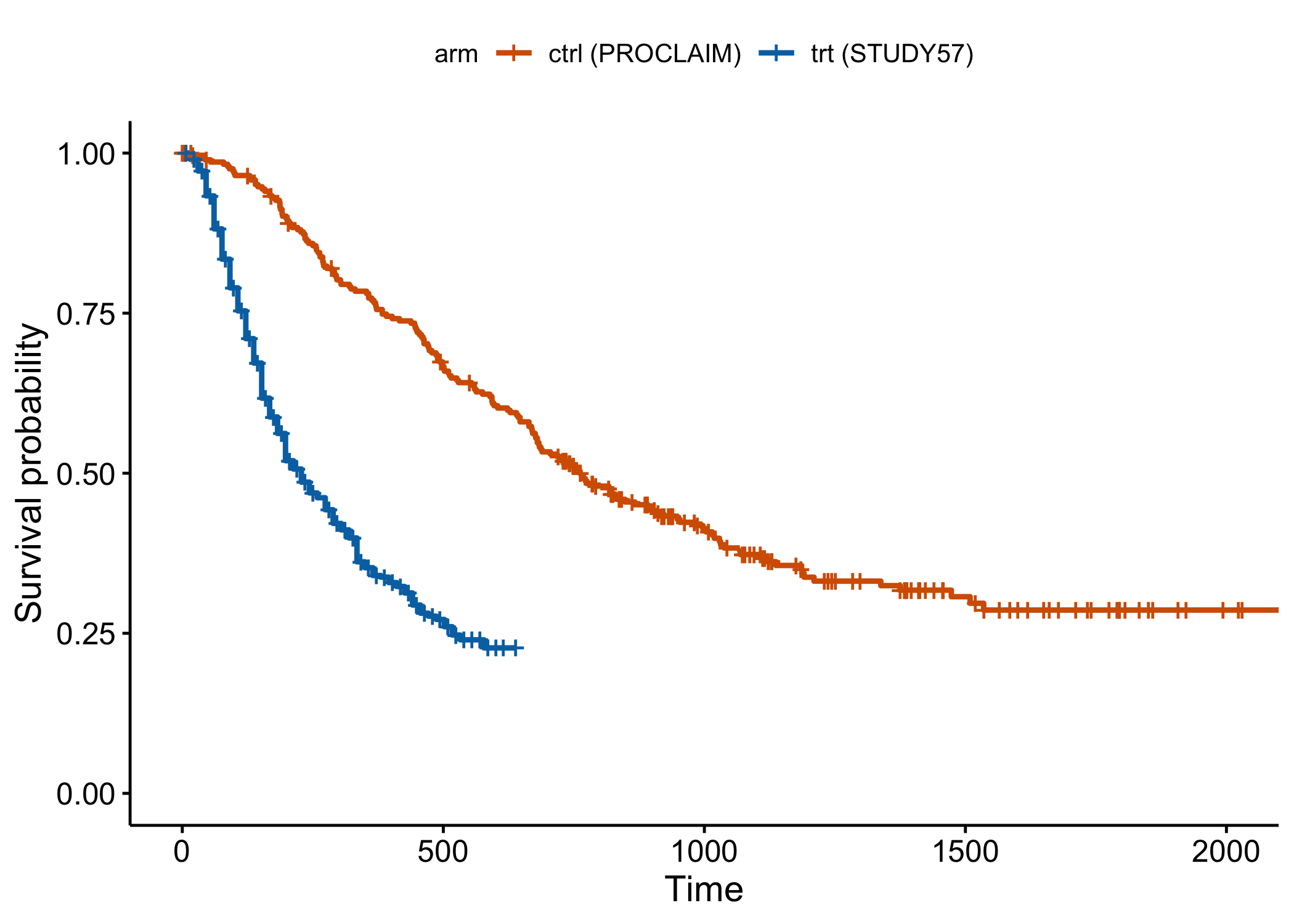}
		\caption{}
		\label{fig:KM13}
	\end{subfigure}
	\begin{subfigure}[b]{0.5\textwidth}
		
		\includegraphics[width=\textwidth]{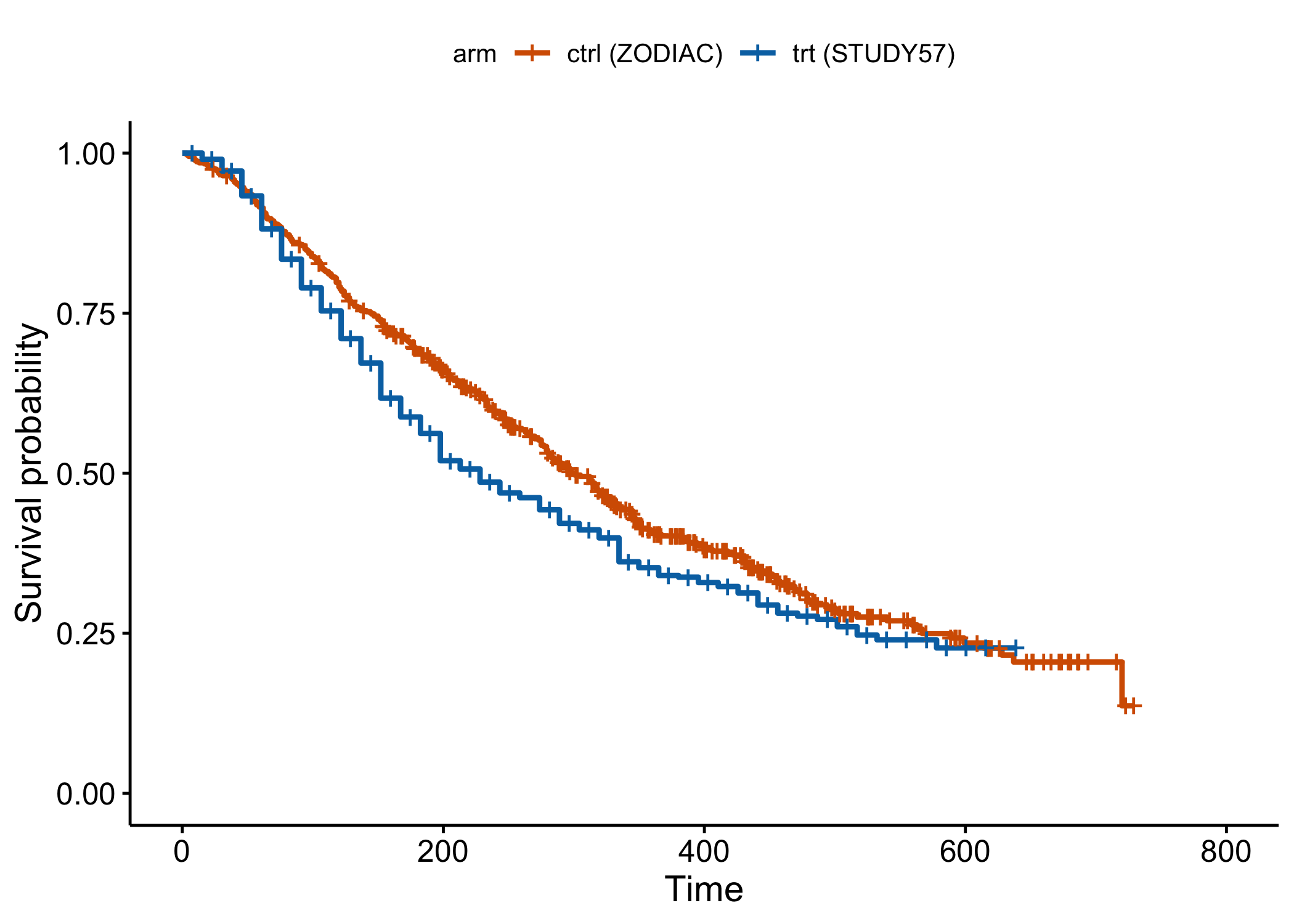}
		\caption{}
		\label{fig:KM14}
	\end{subfigure}\\
	\caption{ {\small Naive comparison of control arms overall survival rate from each trial against the treatment arm data of STUDY57 -- KM curves for STUDY57 using control arm data from (a) STUDY57, (b) INTEREST, (C) PROCLAIM, and (d) ZODIAC.}}
	\label{fig:KM}
\end{figure}

A naive approach is to use the data from the control arms of all four trials in an analysis against the treatment arm of STUDY57. A glance at Figure~\ref{fig:KM} reveals that the results of such an analysis, owing to including an unexchangeable patient population, would be negatively biased with a negative effect estimate for the treatment on patient survival. The goal is, therefore, to incorporate control data of a subset of  individual patients who could plausibly belong to the target study population based on their characteristics and survival outcome with an adequate weight. The methods described in the following section are proposed to achieve this goal. We will revisit this example in Section~\ref{Sec:app} where we apply the proposed methodology and make comparisons.

\section{Methodology}\label{Sec:methods}
While the focus of the present work is use of historical controls for analysis of clinical trials, the proposed method can be generally applied to Bayesian inference. Therefore, for the sake of brevity, in the notation used in this section we will not make  a distinction between control and treatment arm data.

Consider a (concurrent) study that is designed to estimate a set of parameters, $\boldsymbol{\theta}$ , based on sample data $\mathcal{S}_c = (S_{1,c}, \ldots, S_{N_c,c})$ where $S_{n,c}$ indicates all the  available data on subject $n$ in the concurrent study and $N_c$ is the concurrent study sample size. Suppose that $H$ historical studies are available whose data may be used to improve the inference. The data of each historical study $h = 1, \ldots, H$ are denoted by $\mathcal{S}_h = (S_{1,h}, \ldots, S_{N_h, h})$ where $N_h$ denotes the size of historical study $h$. 

Bayesian inference may be performed using only the concurrent study data via the posterior distribution of the parameters given the data,
\vspace{-.2cm}
\begin{equation*}
\pi_{NP}(\boldsymbol{\theta}\mid \mathcal{S}_c) \propto \pi_0(\boldsymbol{\theta}) \pi(\mathcal{S}_c \mid \boldsymbol{\theta}),
\end{equation*}
where $\pi_0(\boldsymbol{\theta})$ is a non-informative prior distribution and $
\pi(\mathcal{S}_c \mid \boldsymbol{\theta}) = \prod_{n = 1}^{N_c} \pi(S_{n,c} \mid \boldsymbol{\theta})
$ is the likelihood.

Alternatively, we may use the historical data with equal weight as that of the present data through the  following informative prior that is based on the likelihood of historical data,
\vspace{-.2cm}
\begin{equation*}
\pi_{FH}(\boldsymbol{\theta})  \propto \pi_0(\theta)\prod_{h = 1}^H\pi(\mathcal{S}_h \mid \boldsymbol{\theta}) =  \pi_0(\theta)\prod_{h = 1}^H\prod_{n = 1}^{N_h}\pi(S_{n,h}\mid \boldsymbol{\theta}).
\end{equation*}
This approach can significantly bias the inference results in presence of study heterogeneity. 

We propose the following individually weighted prior motivated by the power priors but assigning powers to individuals rather than studies,
\vspace{-.2cm}
\begin{equation}
\label{eq:IPD_w}
\pi_{IW}(\boldsymbol{\theta})  \propto  \pi_0(\theta)\prod_{h = 1}^H\prod_{n = 1}^{N_h}\pi(S_{n,h}\mid \boldsymbol{\theta})^{\omega_{n,h}}.
\end{equation}
The power $\omega_{n,h}$ are specified such that subjects who are considered ``eligible" under the concurrent study population will receive a larger weight in the likelihood. 

While $\pi_{IW}(\boldsymbol{\theta})$ moderates the amount of information contained in the prior, there is risk of overpowering the likelihood.  A large number of individual patient data with small weights can result in sufficient information to bias the inference. Therefore, we propose to use only the portion of the IPD with corresponding weights above a specific threshold,
\vspace{-.2cm}
\begin{equation}
\label{eq:twp}
\pi_{TIW}(\boldsymbol{\theta})  \propto  \pi_0(\boldsymbol{\theta})\prod_{h = 1}^H\prod_{n = 1}^{N_h}\pi(S_{n,h}\mid \boldsymbol{\theta})^{\omega_{n,h}\mathbbm{1}(\omega_{n,h}>\rho)}.
\end{equation}
The main challenge is to define the power $\omega_{n,h}$ such that they meaningfully represent eligibility under the target study population. \cite{StuColBra01} used propensity scores to measure generalizability of clinical trial results to the target population. However, propensity scores do not capture complex data structure including non-linear relationships between covariates and outcomes. In the following we address this issue and provide an intuitive approach for specifying the truncation threshold, $\rho$.

\subsection{Specification of the weights} \label{sec:weights}
We propose two methods for specifying the individual weights in the likelihood according to the types of available data. The first method is based on the distance of each individual to the target population that is estimated by the Mahalonobis distance of the individual to the concurrent study distribution. The Mahalanobis distance is a simple and reliable dissimilarity measure as long as all the variables are continuous and their joint distribution can be characterized by a mean vector and a covariance matrix. The second method is based on a \emph{similarity model} that is defined to capture the joint distribution of variables within the concurrent study when the Mahalanobis distance is not an appropriate measure due to presence of discrete variables or highly skewed or multimodal distributions. The weights can then be specified as either the posterior predictive probability of each historical patient data given the present data or their likelihood under the similarity model whose parameter point estimates are obtained from the current study data. 

\subsubsection{Mahalanobis distance}
The powers $\omega_{n,h}$ should be specified such that subjects who better fit the target study population receive larger weights. Considering the concurrent study as the most representative sample of the target population, the weight of every patient in historical studies is specified as a function of their Mahalanobis distance to the concurrent study sample. The Mahalanobis distance is defined based on the joint distribution of all the common variable among studies, i.e, response and covariates, characterized by a mean vector and a covariance matrix. Calculating the Mahalanobis distance does not require the normality assumption. However, as the reference distribution departs from an elliptical distribution that is well characterized by its sample mean and covariance matrix the Mahalanobis distance becomes less appropriate a measure of the membership of a data point to the reference distribution.

Let $S_{n,h}$ denote the vector of all (continuous) variables including covariates and outcomes for patient $n$ in historical study $h$. The distance of patient $n,h$ to the target population is estimated as 
\vspace{-.2cm}
\begin{equation*}
d_{n,h} = \sqrt{(S_{n,h} - \boldsymbol{\mu})^T\boldsymbol{\Sigma}^{-1}(S_{n,h} - \boldsymbol{\mu})},
\end{equation*}
where $\boldsymbol{\mu}$ and $\boldsymbol{\Sigma}$ are the sample mean vector and covariance matrix of the concurrent study.

The weights are then obtained as follows
\vspace{-.5cm}
\begin{equation*}
\omega_{n,h} = 1 - \mathcal{G}(d_{n,h}),
 \vspace{-.5cm}
\end{equation*} 
where $\mathcal{G}$ maps $d$ monotonically onto $(0,1)$,
\vspace{-.2cm}
\begin{equation}
\label{eqn:transform}
\mathcal{G}(d_{n,h}) = \frac{d_{n,h}-\min_{n}d_{n,h}}{\max_{n}d_{n,h} - \min_{n}d_{n,h}}.
 \vspace{-.5cm}
\end{equation}
Note that the minimum and maximum of the Mahalanobis distances used in the mapping are taken across all calculated distances for all the available data including those of the concurrent study. 

With this definition using the information of subjects whose weights are larger than a given threshold $\omega_0$ is equivalent to selecting the subjects whose distance to the target population is within a certain threshold,
 \vspace{-.5cm}
\begin{equation*}\omega_{n,h} > \omega_0  \iff  d_{n,h}< \mathcal{G}^{-1} (1 - \omega_0) \doteq \delta_0,
 \vspace{-.5cm}
\end{equation*} 
where $\delta_0$ can be specified as a quantile of the distribution of distances within the concurrent study. For example if $\delta_0$ is the 95\% quantile of the distance distributions, any historical individual patient data that demonstrates characteristics that fall outside the centre 95\% of the concurrent study data distribution are excluded from the prior.

\subsubsection{Similarity model}
\label{sec:simModel}
The Mahalanobis distance is not appropriate as a dissimilarity measure when discrete variables are present or when the joint distribution of data is not well-characterized by a single mean vector and covariance matrix. Examples of such cases are multimodal and highly skewed distributions. Generalizations of the Mahalanobis distance have been proposed in the literature for mixed discrete and continuous variables \citep{BarDau95, BedLapPow2000, LeoCar05}. The generalized distances are based on models over the joint distribution of variables, namely modeling nominal variables according to a multinomial distribution and assuming multivariate Gaussian distributions for the continuous variables under each level (or level combination) of the nominal variables. 

Similarly, we propose a model-based approach for cases with a mix of discrete and continuous variables. However, instead of defining a distance measure that needs to be converted into a similarity measure, we use the propensity of the historical data given by either the posterior predictive probability in a fully Bayesian approach or its likelihood under the similarity model with parameters estimates obtained from the concurrent data. 

Note that the similarity model is not necessarily the same as the analysis model in that all the variables (including covariates) are assumed to follow a probability distribution. The reason is that joint modelling of all the existing variables is crucial for calculating a similarity measure that reflects patient/study differences while in most clinical trial data analysis, covariates are treated as fixed. Moreover, the similarity model does not have to provide the best fit to the data and therefore can be simpler than the analysis model. For example, one could model all continuous variables as Gaussian random variables for measuring similarity despite presence of mild non-Gaussianity in the data.

Consider the joint similarity model of a set of variables represented by $\mathcal{S}$, denoted by
 \vspace{-.3cm}
\begin{equation}
\label{eq:similarity_model}
\pi(\mathcal{S}\mid \boldsymbol{\psi}),
 \vspace{-.3cm}
\end{equation}
where $\boldsymbol{\psi}$ is the vector of model parameters. As noted above, $\boldsymbol{\psi}$ is generally not identical to $\boldsymbol{\theta}$ since the similarity model is not the same as the analysis model. However, $\boldsymbol{\psi}$ and $\boldsymbol{\theta}$ may have common components.

The weight $\omega_{n,h}$ can then be obtained as the posterior predictive density of patient $n$ in study $h$ given the concurrent study data,
 \vspace{-.5cm}
\begin{equation*}
\label{eq:postpred}
\pi(S_{n,h}\mid \mathcal{S}_c) = \int \pi(S_{n,h}\mid \boldsymbol{\psi}) \pi(\boldsymbol{\psi}\mid \mathcal{S}_c)d\boldsymbol{\psi},
 \vspace{-.5cm}
\end{equation*}

   \begin{equation}
 \label{eq:ML_omega}
 \hat{ \omega}_{n,h}= \mathcal{G}(\hat{ \pi}(S_{n,h}\mid \mathcal{S}_c)).
 \end{equation}
 where $\mathcal{G}$ is given in (\ref{eqn:transform}).
 
 Alternatively, taking an empirical Bayes approach, the likelihood of historical data given estimates $\hat{\psi}$ may be used. 
 \begin{equation}
 \label{eq:ML}
\hat{ \omega}_{n,h}= \mathcal{G}( \pi(S_{n,h}\mid \hat{\psi})).
 \vspace{-.5cm}
 \end{equation}
 
  Similar to what was explained for the Mahalanobis distance method, the truncation threshold is obtained as a quantile of the weights in the concurrent study that are calculated analogously based on the posterior predictive density or likelihood of every patient within the concurrent study.

\section{Simulation study}\label{Sec:sim}
In this section we make comparisons between the proposed IPD-weighted prior and two of the existing approaches for borrowing historical information that were discussed in Section \ref{sec:intro} namely power priors and meta-analytic predictive priors. The data for the simulation study are generated for fictitious clinical trial with a continuous normally distributed outcome that depends on a measured covariate and the treatment with effect size $\theta = 0.5$. Historical control data are generated under five main scenarios: 1)  exchangeable: the concurrent and historical control data follow the same distributions for the measured covariate and outcome; 2) Partially exchangeable 1: the measured covariate in a portion of the historical study has a different distribution than that of the concurrent study; 3) Partially exchangeable 2: the outcome distribution in a portion of the historical data is different than that of the concurrent data due an unmeasured effect; 4) Unexchangeable 1: the measured covariate in the historical control has a different distribution than that of the concurrent control; 5) Unexchangeable 2: the outcome distribution in the historical data is different than that of the concurrent data due to an unmeasured effect larger than 3 times the outcome standard deviation.  6) Unexchangeable 3: the outcome distribution in the historical data is different than that of the concurrent data due to an unmeasured effect of size 1 standard deviation. 

Three sample sizes are explored for the concurrent study, $N_c = 25, 50, 100$ while the size of historical control is held fixed at $N_h=100$ in all simulation scenarios.  The measured covariate and the outcomes are generated from Gaussian distributions as follows,
\begin{equation*}
\mathbf{x_c}\sim \mathcal{N}(1, 1) \hskip 20pt \mathbf{y_c}\mid \mathbf{x_c} \sim \mathcal{N}(\beta \mathbf{x_c} + \theta \mathbf{z}, 1),
\end{equation*}
where $z$ is the treatment assignment vector generated from a binomial distribution with equal allocation probability. 

The historical control data are generated from
\begin{equation*}
\mathbf{x_h}\sim \mathcal{N}(\mu_{xh}, 1) \hskip 20pt \mathbf{y_h}\mid \mathbf{x_h} \sim \mathcal{N}(\delta_h + \beta \mathbf{x_h}, 1),
\end{equation*}
The parameter $\delta_h$ is a constant that represents a shift in the outcome distribution. The parameters of the historical control distributions for various simulation scenarios are given in Table~\ref{table1}.

\begin{table}
	\caption{Parameter values for the historical study distribution for various simulation scenarios}
	\centering
	\begin{tabular}{c|c|c}
	 Scenario& $\mu_{xh}$& $\delta_h$\\
	\hline
	exchangeable &  1& 0 \\
	partially exchangeable 1 & 1 or 6 (mixture)  & 0\\
	partially exchangeable 2 & 1 &   0 or 5 (mixture)\\
	unexchangeable 1 &  6&   0\\
	unexchangeable 2 &   1&   5\\
	unexchangeable 3 &   1&   1\\
	\end{tabular}

\label{table1}
\end{table}

The simulated data are analysed in the Bayesian framework using priors that are constructed by the proposed method and the other existing methods as described below. The analysis model is given by,
 \vspace{-.5cm}
\begin{align}
\label{eqn:general_model}
\pi(\theta, \beta_0, \beta, \sigma^2 \mid \mathbf{y}) \propto \pi_0(\theta, \beta_0, \beta, \sigma^2)&\prod_{n = 1}^{N_h}\phi(y_{n}\mid \beta_0+\beta x_{n}, \sigma^2)^{\omega_{n}}\nonumber\\ &\prod_{n = 1}^{N_c}\phi(y_n\mid \beta_0 + \beta x_n + \theta z_n, \sigma^2),
 \vspace{-.5cm}
\end{align}
where $\pi_0(\theta, \beta_0, \beta, \sigma^2)$ is an independent non-informative prior and $\phi(.\mid a, b^2)$ denotes the Gaussian probability density function with mean $a$ and variance $b^2$. The weights $\omega_{n}$ are given for each of the competing methods for prior construction as follows,

	 {\bf No prior:} The historical controls are excluded from the analysis, i.e., $\omega_{n}=0, \hskip 15pt \forall n$;
	 
	{\bf Full history prior:} The historical controls are fully combined with the concurrent study controls, i.e., $\omega_{n}=1, \hskip 15pt \forall n$;
	 
	  {\bf Power  prior:} Each historical study is assigned a weight that is obtained as the penalized likelihood-type criterion proposed by \cite{IbrCheSin03},
	   \vspace{-.5cm}
	  $$\omega_{n} = \omega^{\text{opt}}.\vspace{-.5cm}$$
     For the normal model used in the simulation study these weights can be obtained analytically given a flat prior $\pi_0(\beta_0, \beta, \theta) \propto 1$ and known $\sigma^2$. However, in general the penalized likelihood-type criterion involves integrating the posterior over model parameters which requires numerical integration in most cases.
	  
	  {\bf Individually weighted prior:} Using historical controls with each individual weighted, i.e., $\omega_{n,h} = 1 - \mathcal{G}(d_{n})$ for $n = 1, \ldots, N_h$.
	  
	  {\bf Truncated individually weighted prior:} Similar to the individually weighted approach but including individuals from the historical study who have a sufficiently large weight;
	  $$\omega_n = \begin{cases} 1 - \mathcal{G}(d_{n}) & \text{if } 1 - \mathcal{G}(d_{n})>\rho \\
	  0 & \text{if } 1 - \mathcal{G}(d_{n})<\rho.
	  \end{cases}
	  $$
	  	  where $d_n$ is the Mahalanobis distance of individual $n$ to the concurrent study sample.
	 
	  {\bf meta-analytic predictive prior:} The meta-analytic predictive prior is the only prior that cannot be described by the general model in (\ref{eqn:general_model}),
	   \vspace{-.5cm}
	 \begin{align*}
	 \pi_{MAP}(\theta, \delta_s, \beta, \sigma^2, \mu_\delta, \tau \mid \mathbf{y}_c, \mathbf{y}_h) \propto \pi_0(\mu_\delta, \tau^2,& \theta, \beta, \sigma^2)\phi(\delta_c, \delta_h \mid \mu_\delta, \tau^2)\nonumber\\
	 &\prod_{n = 1}^{N_h}\phi(y_{n}\mid \delta_h+\beta x_{n}, \sigma^2)\nonumber\\ &\prod_{n = 1}^{N_c}\phi(y_n\mid \delta_c + \beta x_n + \theta z_n, \sigma^2), \vspace{-.5cm}
	 \end{align*}
	 where $\mu_\delta$ and $\tau^2$ are the common mean and variance of the study random effects. The variance $\tau^2$ is the key parameter that controls the amount of information borrowed according to the study heterogeneity. Note that the meta-analytic predictive prior is in fact the ``correct" model as it captures the data generating model.

The methods are compared in terms of frequentist power, i.e., the probability that the lower bound of 95\% credible intervals for $\theta$ are greater than zero, estimated over the sampling distribution; the root mean squared error computed by the square root of the mean squared errors between posterior draws of $\theta$ and the true value; bias computed as the deviation of posterior mean from the true value; and the width of the 95\% credible intervals that represent the posterior uncertainty. 

\begin{figure}[t]

			\centering
			\includegraphics[width=1\textwidth]{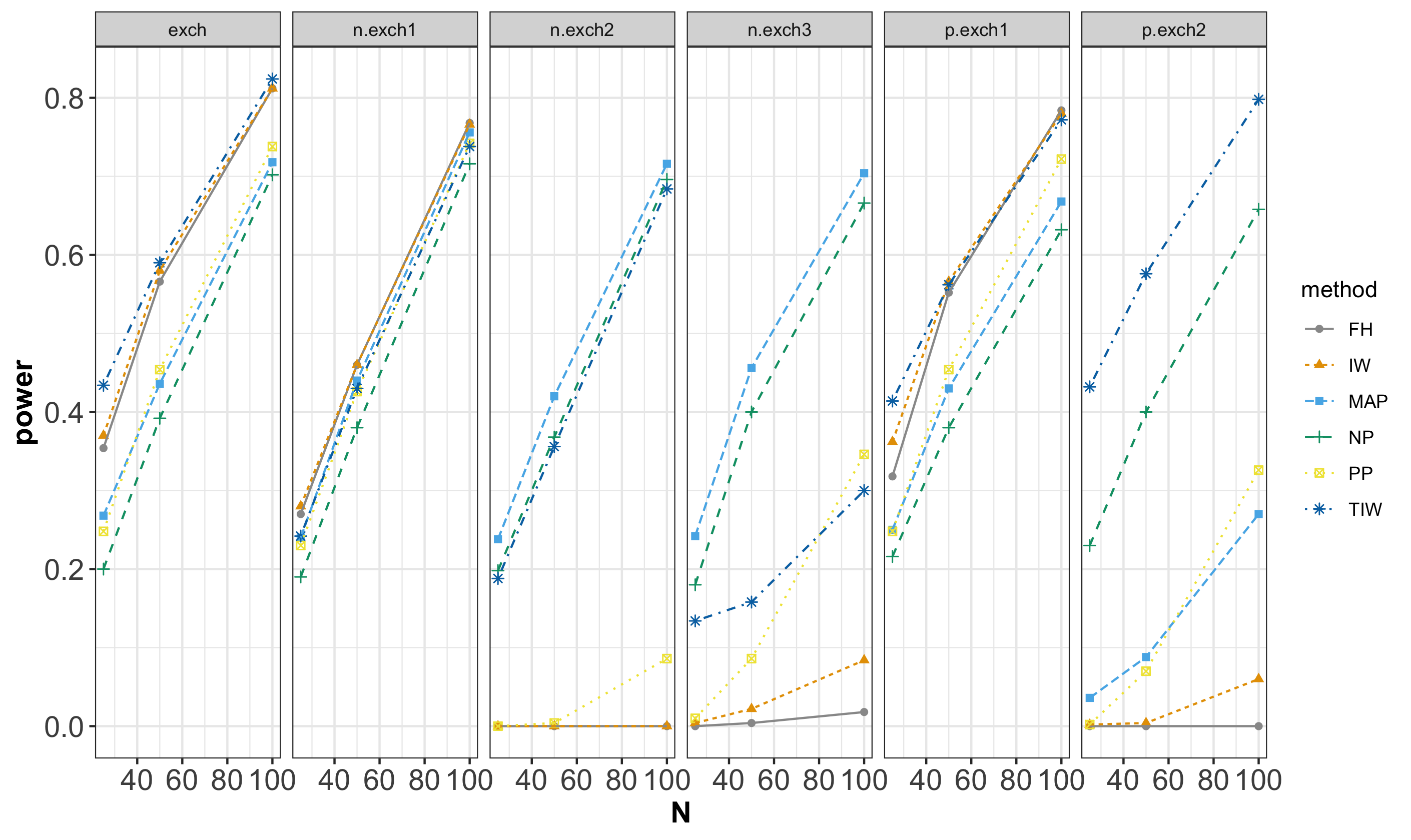}
			\caption{Power for detecting a positive effect for six simulation scenarios (column panels), six methods incorporating various amounts of external control data (legend) and increasing sample sizes (X axis).}
			\label{fig:power}
		\end{figure}

\begin{figure}[t]	

		\centering
		\includegraphics[width=1\textwidth]{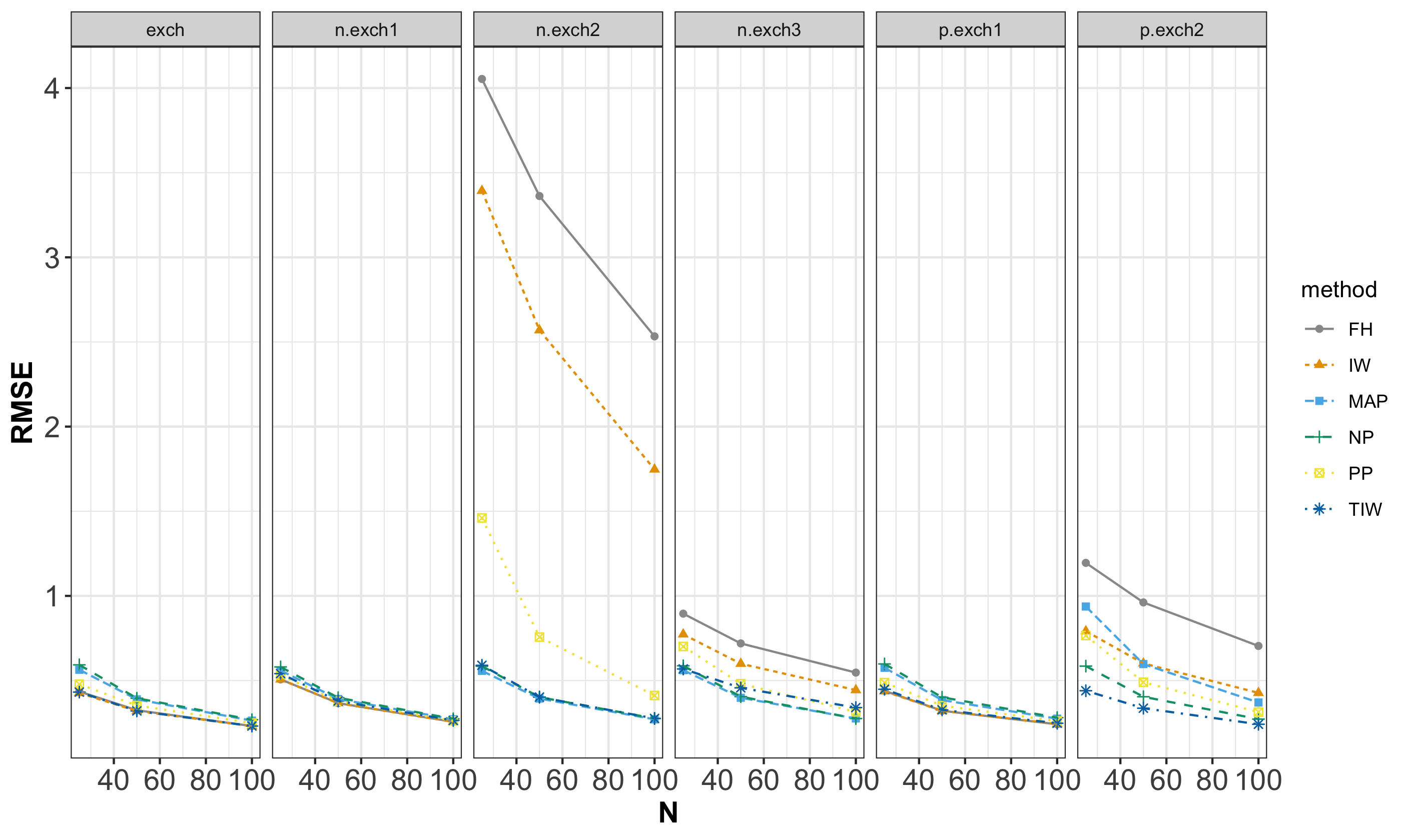}
		\caption{RMSE for estimating the treatment effect averaged over 500 simulation iterations for six simulation scenarios (column panels), six methods incorporating various amounts of external control data (legend) and increasing sample sizes (X axis).}
		\label{fig:RMSE}
	\end{figure}

\begin{figure}[t]	

		\centering
		\includegraphics[width=\textwidth]{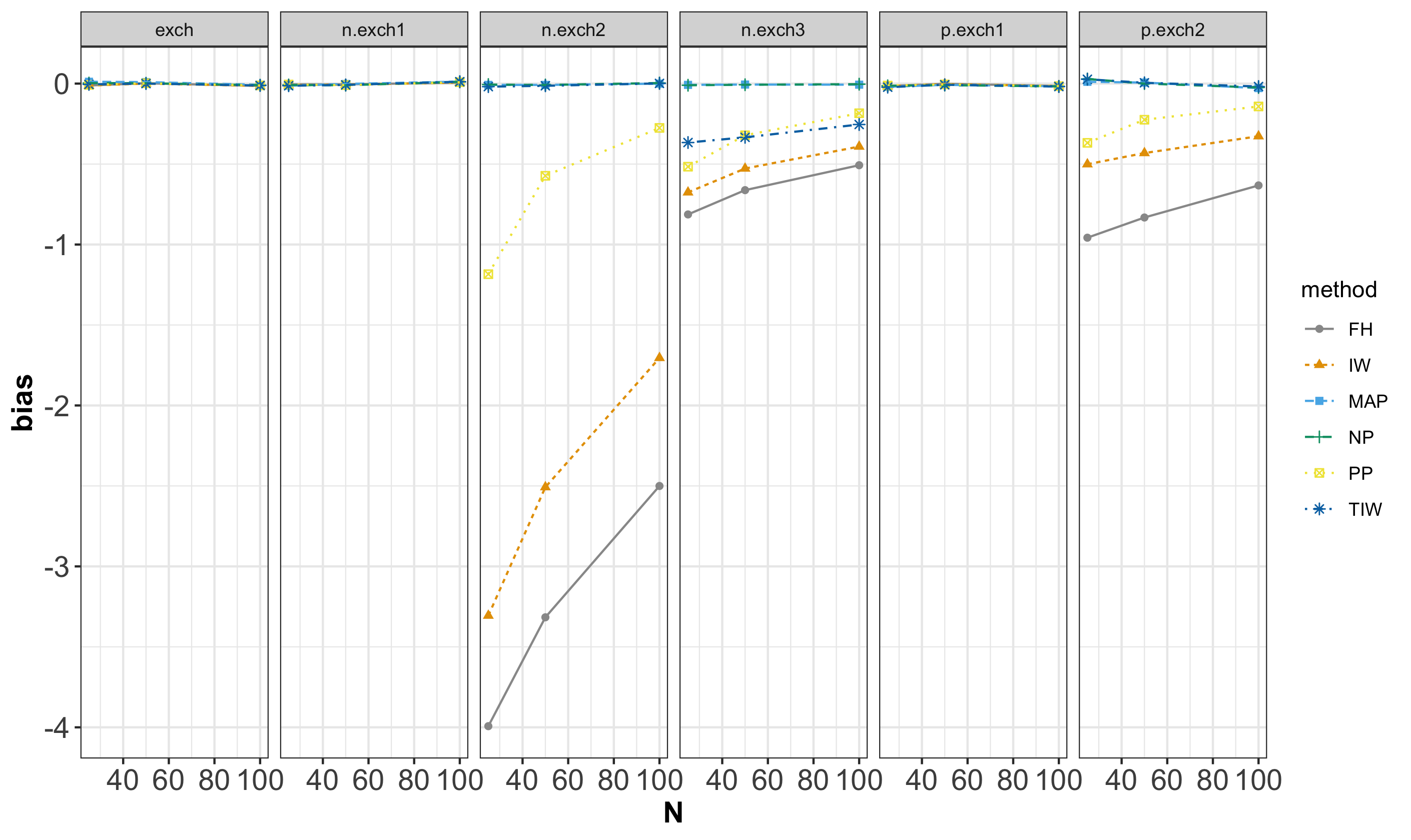}
		\caption{Bias for estimating the treatment effect averaged over 500 simulation iterations for six simulation scenarios (column panels), six methods incorporating various amounts of external control data (legend) and increasing sample sizes (X axis).}
		\label{fig:bias}
	\end{figure}

\begin{figure}[t]	

	\centering
	\includegraphics[width=\textwidth]{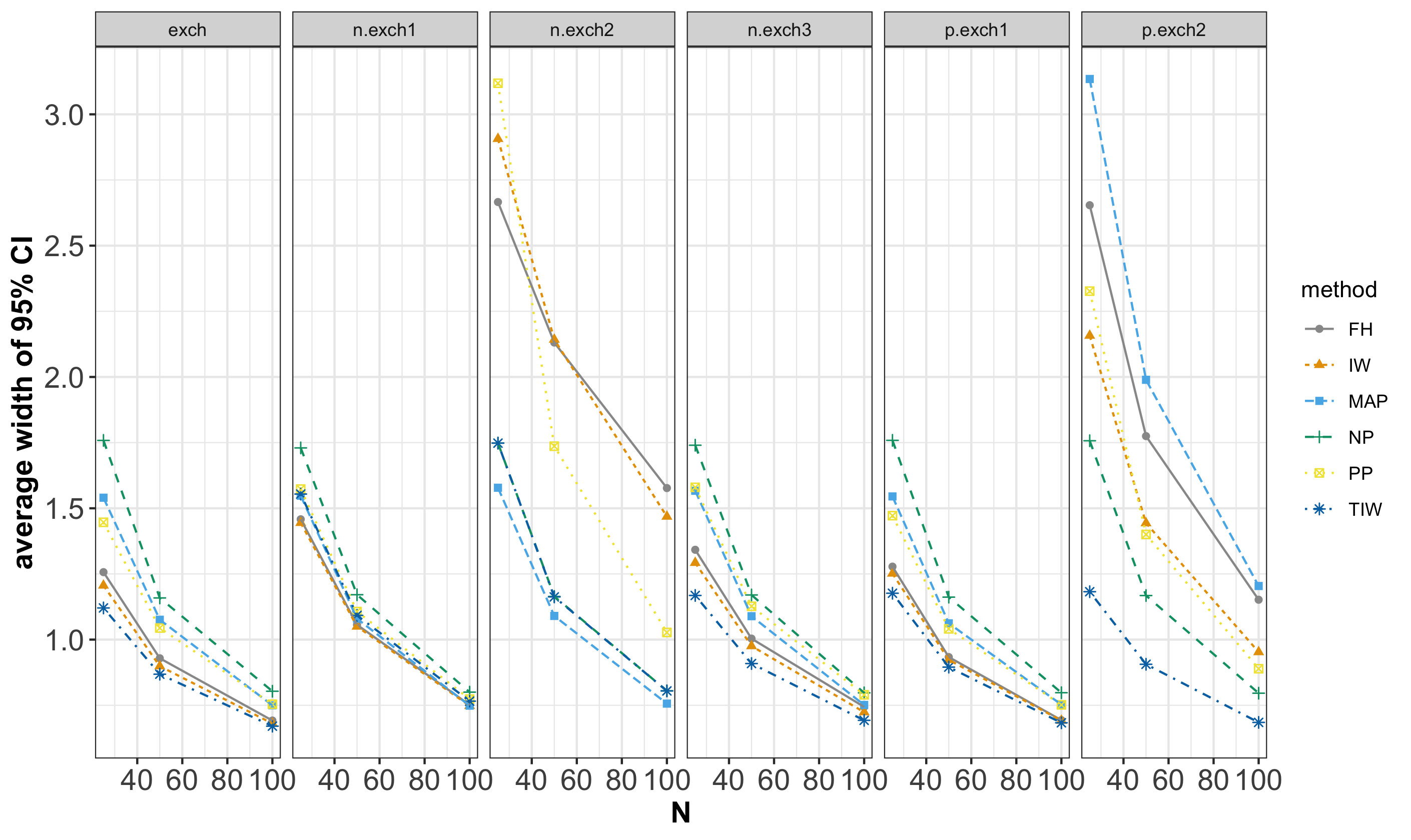}
	\caption{Width of 95\% credible intervals for effect size estimates, averaged over 500 simulation iterations for six simulation scenarios (column panels), six methods incorporating various amounts of external control data (legend) and increasing sample sizes (X axis).}
	\label{fig:IQE}
\end{figure}

Figures~\ref{fig:power}-\ref{fig:IQE} show the simulation results. The highest gain in power is achieved by the proposed approach (TIW) under the partially exchangeable 2 scenario (last panel). When non-exchangeability or partial exchangeability is due to differences in measured covariates all methods achieve a reasonable level of power with those that leverage the maximum amount of data (FH, IW and TIW) leading. However, when differences in outcome distributions are due to a study related effect that is not measured through the observed covariates only TIW is able to maintain satisfactory power followed by NP. The loss in power among FH, IW and power priors (PP) is due to bias (Figure \ref{fig:bias}). This bias is due to incorporating the unexchangeable portion of the data with weight 1 in FH that results in the largest bias and with smaller weight in IW that results in an improvement over FH but is still negatively biased. Even the power prior results in a small negative bias since the same weight value is assigned to the historical data and the unexchangeable portion is not discarded. This bias however, becomes smaller as the sample size of the current study increases and the priors' influence diminishes. MAP yields unbiased estimates under all scenarios since it assigns separate parameters to the concurrent and historical control means. However, in presence of any amount of unexchangeability the study heterogeneity results in a large variance ($\tau^2$)  estimate that in turn increases the posterior uncertainty (Figure~\ref{fig:IQE}) and reduced power.

The non-exchangeable 3 scenario is an interesting case (4th panel from left in Figures ~\ref{fig:power}-\ref{fig:IQE}). This is the case where the outcome distribution within the historical control data are shifted by about one standard deviation of the concurrent outcome distribution. The historical control data will therefore receive large weights under the TIW which result in negatively biased estimates for the treatment effect and therefore a loss in power. In fact, the only two methods that are able to provide unbiased estimates are NP that excludes all historical data and MAP that allows separate parameter specification. Such a scenario requires more careful investigation: whether to include the subset of subjects whose outcome distribution is well contained the current study outcome distribution but may exhibit different distributional characteristics that change the overall shape of the control outcome distribution should be addressed within the context. While the change in control outcome distribution can result in biased estimates for the ``true" treatment effect for the population of patients represented by the current study data, it identifies the subgroup of individuals whose inclusion shifts the treatment effect estimates. Such a result cannot necessarily be obtained from subgroup analyses within the current study data due to sparsity of data in subgroups, neither would it be possible through separate analyses of historical data. Curiously, this appears to be the case in the NSCLC example that we will discuss in the following section.

\section{Analysis of the NSCLC data}\label{Sec:app}

In this section the proposed methodology is applied to the data introduced in Section~\ref{Sec:me}. STUDY57 is considered as the concurrent trial and the other three studies are used to enrich the control arm data. As mentioned earlier, one of the three historical trials (PROCLAIM) showed significantly different patient and survival characteristics. \cite{DroGolHsu19} showcase analysis of these studies using the meta-analytic predictive prior approach with different levels of borrowing arising from including/excluding PROCLAIM among the historical studies. Given the clear differences in eligibility criteria and control arm definition, this trial should be excluded from the set of historical studies.  However, we choose to include PROCLAIM in our analyses for illustrative purposes. 

Since the data include a number of discrete covariates such as sex and race, for specification of the weights in the individually weighted prior, we use the model-based approach described in \ref{sec:simModel}. In the following, we introduce the similarity model as well as the model used for Bayesian analysis of the STUDY57 trial.

The similarity model is used only to model the control arm data. The outcome is time of death for observed events and the time of lost to follow-up for patients with censored data. Available covariates among all four studies include age (treated as continuous), sex (dichotomous) and race with four categories that was reconstructed as three dummy variables. Let us denote the categorical variable $C$ whose $K$ categories result form level combinations of all categorical variables, i.e., sex, race and the censoring variable. The category of $C$ to which patient $n$ under the control arm of the concurrent trial belongs is denoted by $C^{ctrl}_n$. Similarly, we denote the continuous variables, i.e., age and time of death/censoring, for patient $n$ in the control arm by $D^{ctrl}_n$. The similarity model is then given by,
\vspace{-.5cm}
\begin{align}
\label{eq:sim_model}
&C^{ctrl}_n \sim \text{Multinom}(1, (p_1, \ldots, p_K)) \nonumber \\
&D^{ctrl}_n \mid C^{ctrl}_n = k  \sim \mathcal{N} (\mu_k, \Sigma_k), \hskip 15pt 
k = 1, \ldots, K
\end{align} 
where $\mu_k$ and $\Sigma_k$ are the mean vector and covariance matrix of continuous variables in category $k = 1, \ldots, K$. Parameters $(p_k, \mu_k, \sigma^2_k)$, $k = 1, \ldots, K$ are estimated from the control data of STUDY57. The weight for a patient in the control arm of any of the other three studies is given as,
\begin{equation*}
\hat{\omega}_{n,h} = \mathcal{G}\left(\sum_{k = 1}^K \hat{p}_k\phi(D_{n,h}\mid \hat{\mu}_k, \hat{\Sigma}_k) I(C_{n,h}=k)\right)
\end{equation*}
where $I(C_{n,h}=k)$ indicates if individual $n$ within the historical control $h = 1, 2, 3$ is in category $k$. 

It is likely that some categories of $C$ contain no or few observations and therefore the corresponding parameters cannot be estimated precisely within the present study control arm. In this case, however, $\hat{p}_k$'s, i.e. the estimated probabilities associated with empty or small categories will be zero or close to zero which results in negligible contribution of historical observations that fall within these categories. Therefore, the quality of parameter estimates within these categories is of little concern.

The analysis model is defined as a Bayesian hierarchical model with the proportional hazards assumption that is used to analyse the survival data (control and intervention arms) of the concurrent study. More specifically the likelihood is given by 
\vspace{-.5cm}
$$\pi(\mathbf{y} \mid \alpha, \lambda_n) = \prod_{n = 1}^{N_C}f(y_n\mid\alpha, \lambda_n)^{\nu_n}S(y_n\mid\alpha, \lambda_n)^{(1-\nu_n)}, \vspace{-.5cm}$$
where $\mathbf{y}$ is the vectors of responses, $f(y_n\mid\alpha, \lambda_n)$ is a Weibull probability density function with shape parameter $\alpha$ and scale parameter $\lambda_n$,
$S(y_n\mid \alpha, \lambda)$ is the Weibull survival function, and $\nu_n = 0$ indicates that patient $n$ is right-censored. The regression model is embedded within the scale parameter,
$$\lambda_n = \delta + \mathbf{x}_n\boldsymbol{\beta} + z_n\theta,$$
where $z_n = 1$ indicates treatment assignment.
The parameter of interest is $\theta$ which represents the treatment effect. The hazard ratio is given as,
\vspace{-.5cm}
$$\text{HR} = \exp{(\theta)}. \vspace{-.5cm}$$
The IPD-based prior is defined as follows,
\vspace{-.5cm}
\begin{equation*}
\pi_{TIW}(\alpha, \boldsymbol{\beta} , \delta, \theta) =\pi_0(\alpha, \boldsymbol{\beta} ,\delta, \theta)\prod_{h=1}^H\prod_{n = 1}^{N_h}[f(y_{n,h}\mid\alpha, \boldsymbol{\beta} , \delta)^{\nu_{n,h}}S(y_{n,h}\mid\alpha,\boldsymbol{\beta} , \delta)^{(1-\nu_{n,h})}]^{\hat{\omega}_{n,h}(1-z_{n,h})}, \vspace{-.5cm}
\end{equation*}
where $\pi_0(\alpha, \boldsymbol{\beta} ,\delta, \theta)$ is an independent uninformative prior. Specifically, $\delta$, $\boldsymbol{\beta}$ and $\theta$ are assigned normal distributions centered at zero with variance $10^6$ and $\alpha$,  is assigned the same normal distribution truncated at zero since $\alpha>0$. The power $(1-z_{n,h})$ indicates that only control arm historical data are incorporated into the prior. Note that the informative prior does not contain information about $\theta$ since it only uses historical data from the control arm.

The posterior kernel is then given as the product of the prior and the likelihood,
\vspace{-.5cm}
\begin{equation*}
\pi(\alpha, \boldsymbol{\beta} , \delta, \theta\mid \mathbf{y}) \propto \pi_{TIW}(\alpha, \boldsymbol{\beta} , \delta, \theta)\pi(\mathbf{y} \mid \alpha, \lambda_n(\boldsymbol{\beta}, \delta, \theta)).
\vspace{-.5cm}
\end{equation*}
Samples are drawn from the above posterior distribution using Markov chain Monte Carlo.

\begin{figure}[t]

		\centering
		\includegraphics[width=\textwidth]{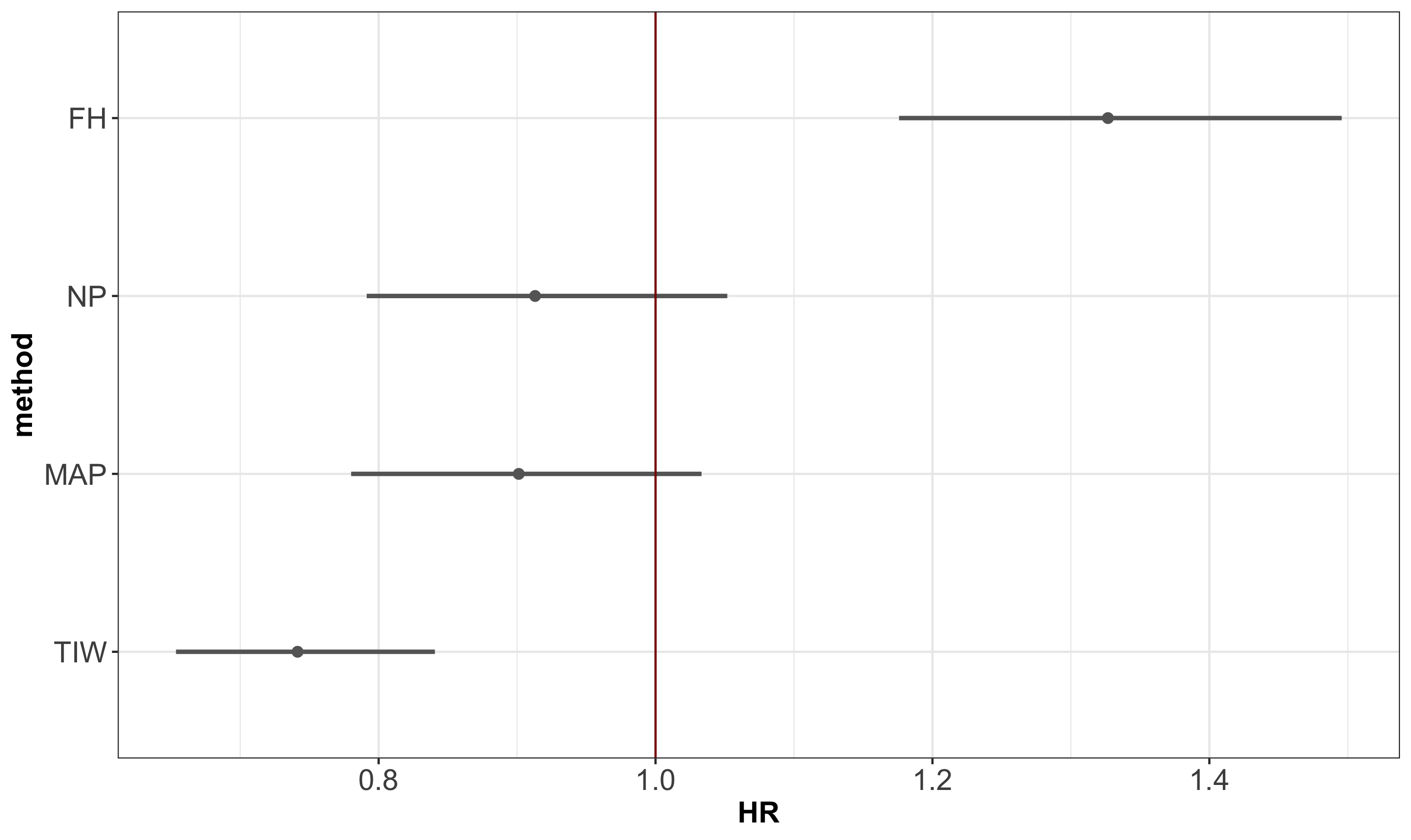}

	\caption{(a) Bayesian 95\% credible intervals for the hazard ratio for STUDY57 obtained by no prior (NP), truncated individually weighted prior (TIW), meta-analytic predictive prior (MAP), power prior (PP) and full historical data (FH)}
	\label{fig:CD_HR}
\end{figure}

\begin{figure}[t]
	\centering
\includegraphics[width=\textwidth]{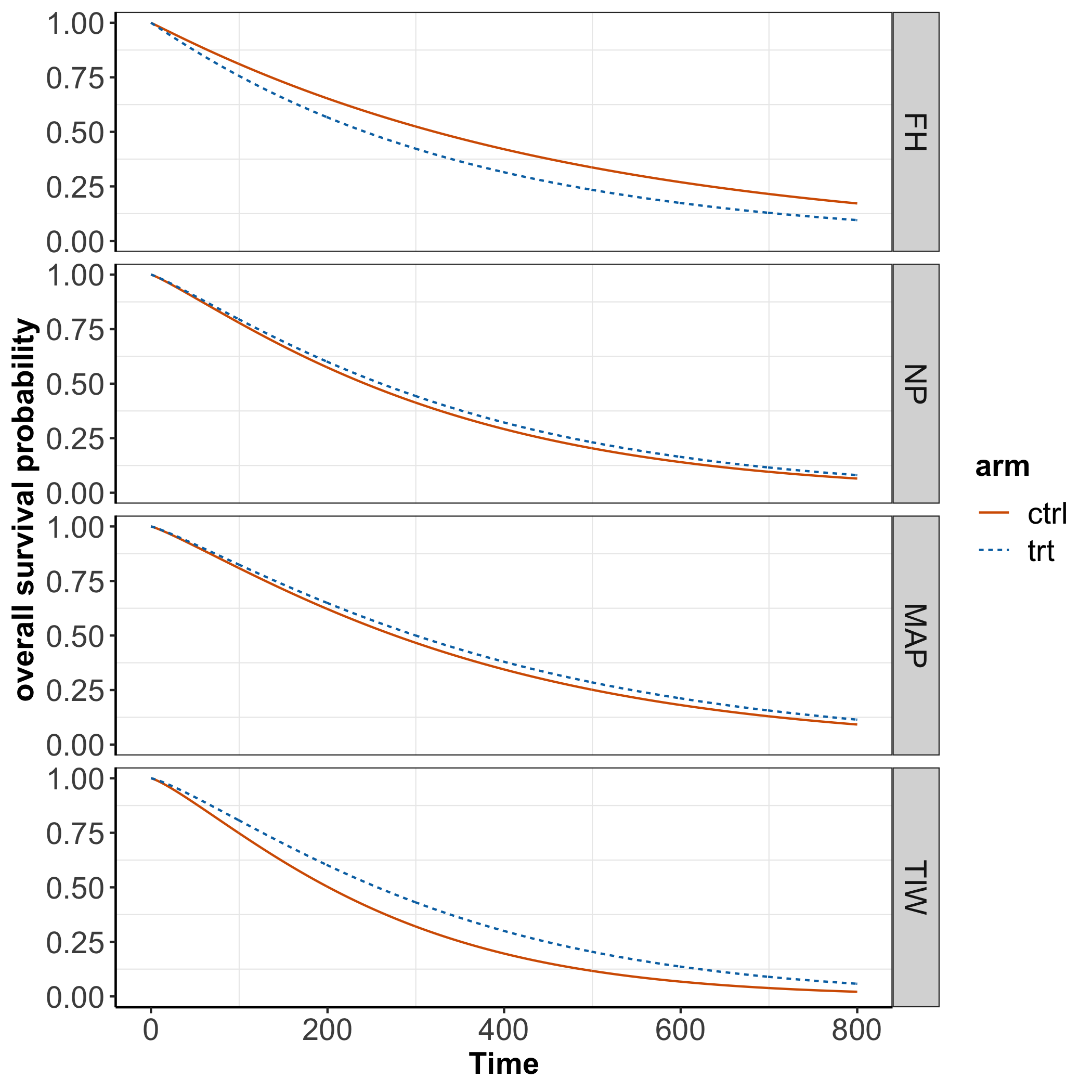}
\caption{Survival probability curves for the control and treatment arms in STUDY57 obtained from the point estimates (posterior means) of the model parameters by each of the four methods.}
\label{fig:SC}
\end{figure}

The data for STUDY57 trial are analysed using the above model with a non-informative prior, the TIW prior, the meta-analytic predictive prior and the naive approach that uses all control data from the other three trials with the same weight as that of STUDY57. Figures~\ref{fig:CD_HR} and \ref{fig:SC} show the 95\% posterior credible intervals of the hazard ration (HR) and the corresponding estimated survival curves for the four methods, respectively. As mentioned earlier, analyses of the data within STUDY57 alone (NP) results in statistically insignificant results with a posterior mean of $HR \approx 0.92$. 

Naively using all the available control data (FH) results in HR estimates that correspond to a negative effect on overall survival. As mentioned earlier, a comparison of the control data from each of the trials against the treatment arm of STUDY57 in Figure~\ref{fig:KM} explains this extreme results. However, considering the differences between patient populations (specially that of PROCLAIM) these estimates are unreliable and the FH analyses is invalid. 

HR estimates and credible intervals obtained by MAP are very close to that of NP. The hierarchical structure of the prior estimates a high level of study heterogeneity which results in negligible level of information borrowing from the historical control data. 

Finally, TIW estimates of HR show a statistically significant effect in the opposite direction of that indicated by FH. By using the subset of historical control data that is most probable under the similarity model (\ref{eq:sim_model}) the effect estimates are shifted to $\hat{HR}\approx 0.75$ with narrower 95\% credible intervals. While these results are compelling, they should not be considered conclusive considering that the selective nature of the proposed priors can result in estimation bias in certain scenarios. Rather, such a result calls for further investigation into the characteristics of the subgroup of control IPD whose inclusion alters the inference and how the inclusion of the select IPD affect the generalizability of the results. Specifically, does the augmented historical control data compliment the concurrent control sample and can be considered a better representation of the target population? Or inclusion of historical control data results in distortion of present study control distribution that was already representative of the target population?
	
\begin{figure}[t]
	\centering
	\includegraphics[width=\textwidth]{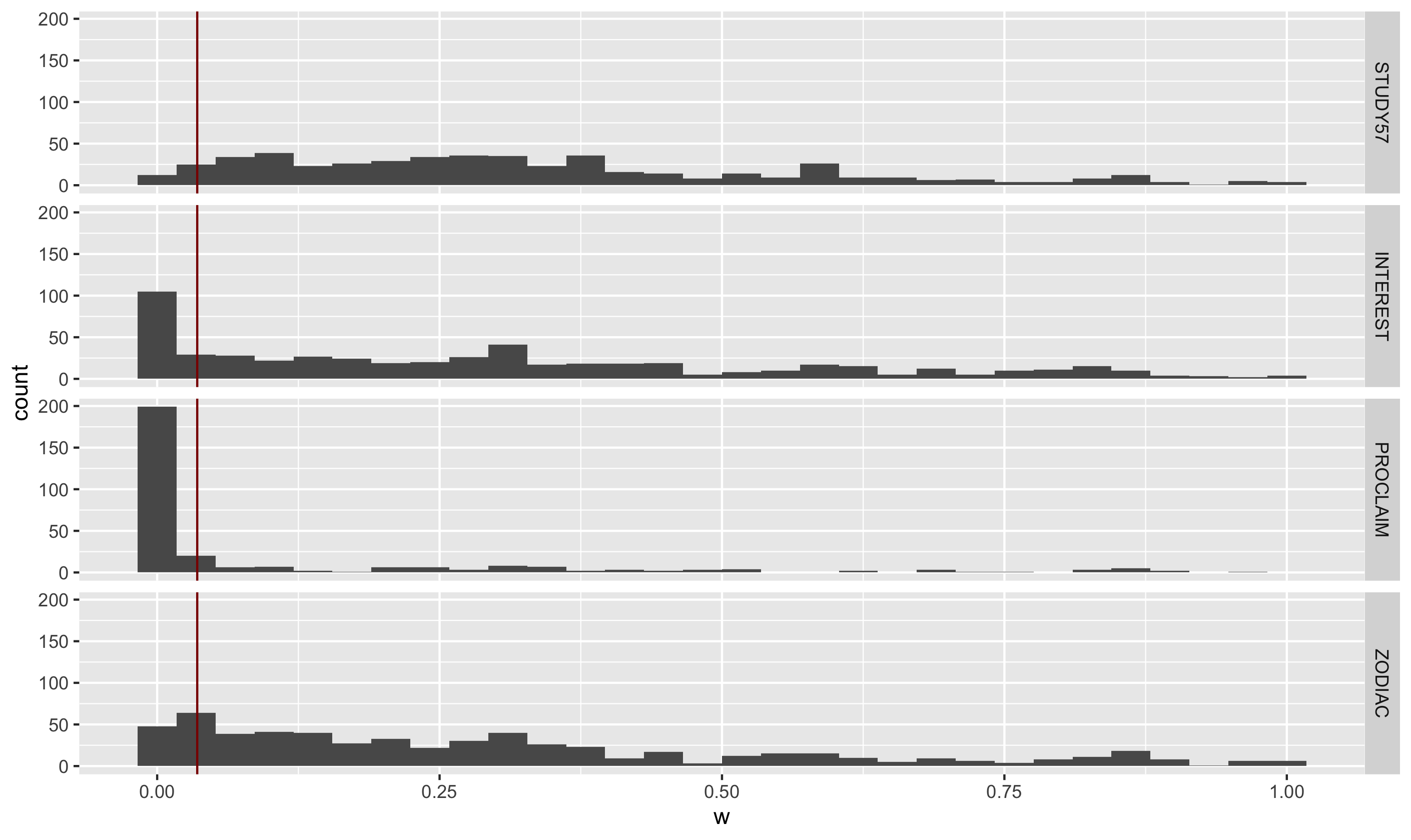}
	\caption{Raw power distribution for the four NSCLC trials. The vertical line shows the lower 5\% quantile of the power distribution for STUDY57. Observations within the other three study whose power is below this threshold receive zero weight under the TIW prior.}
	\label{fig:weights}
\end{figure}

Considering the between study differences and specially the fact that PROCLAIM stands out among the three historical trials as one that should not be used to inform inference, it is interesting to see what percentage of individual patient data from each historical trial is incorporated into the prior. Figure~\ref{fig:weights} shows the distribution of raw powers for the four studies. The vertical line shows the truncation point which is specified as the 5\% quantile of the power distribution for STUDY57, i.e., 95\% of individuals within the control arm of STUDY57 trial have powers greater than this threshold. Note that all individuals within STUDY57 will be included in the analysis model with power one. For the other three trials, INTEREST and ZODIAC contribute to the prior with about 30\% of IPD receiving powers above the cut-off value while from PROCLAIM only 10\% of IPD receive non-zero weights in the prior. 

\section{Discussion}\label{Sec:dis}
In this article we have proposed methodology for incorporating individual patient data from external data sources, such as past studies, to the analysis of clinical trial data with the goal of improving statistical inference. The proposed family of priors can be considered a generalization of power priors where instead of assigning a power to studies, each individual within the historical studies receives a power. The weight or power assigned to each individual is a function of their similarity to the concurrent study population. The similarity is measured through a set of common variables including covariates and outcome(s). The Mahalanobis distance for continuous data and a general model-based approach that suits any data type are recommended for specification of the weights. 

The general weight specification approach relies on a similarity model that is intended to capture important data structure including correlations among variables. We emphasize that the similarity model is not necessarily identical to the analysis model. It can be more complex in that it assigns probability distributions to covariates that may be considered fixed under the analysis model. But it can be simpler in that approximate Gaussian distributions may be used even when data are not entirely Gaussian.

An essential component of the proposed prior construction approach is a cut-off value for the powers and discarding individual patient data whose powers fall below this cut-off value. The truncation is introduced to assure that the prior is constructed based on historical data quality (represented by the weights) rather than quantity (large number of observations with small weights). The intuitive explanation is that any individual data from the historical studies that fall outside a pre-specified credible set of the concurrent study distribution should not be used to inform inference.

%In our simulation studies we did not observe any significant difference in the results from using a similarity model that is exactly the data generating model versus one that assumes an approximate Gaussian distribution for all continuous variables. However, as non-Gaussian features become more dominant in the data, a similarity model that captures these features can result in more appropriate information borrowing.

An outstanding result from the simulation study is the scenario where the past data, although different in distribution, is credible under the present data distribution and therefore receive large weights in the analysis. In this case the individually weighted prior results in estimates that are biased for the effect size parameter assumed for the concurrent data. Curiously, this appears to be the case for the NSCLC data as the proposed prior results in smaller HR and statistically significant effect estimates. As emphasized earlier, this potential bias does not deem the individually weighted priors entirely unreliable but can be used to gain insight towards the effect within patient subgroups. In the NSCLC application, for example, further investigation should be made to understand the characteristics and study membership of the individual patients whose inclusion of outcomes results in a compelling effect, a result that may not necessarily be obtained from a subgroup analysis due to lack of power.

A question that remains open is how to select the variables to be included in the similarity model. This is not a major issue in the NSCLC application included in the present work since the covariates consist of a small number of demographic variables that can be reasonably used to define the study population. However, in cases where there are a large number of shared covariates among studies it is important to select a subset of variables that meaningfully characterize the target population. 

\section*{Data and Code}

The IPD under the control arm for the four NSCLC trials was obtained from Project Data Sphere (\href{http://www.projectdatasphere.com}{http://www.projectdatasphere.com}), an open-source repository of individual-level patient data from oncology trials. No IPD was available under the intervention arms of either trials. Therefore, we recovered IPD for the intervention arm of ZODIAC by digitizing the Kaplan-Meier curves provided in the publication. 

All the Bayesian computation, i.e., posterior sampling, was performed in RStan. The Stan models together with the R script that can be used to reproduce the results of the paper are provided at \href{https://github.com/sgolchi/IPD_prior}{https://github.com/sgolchi/IPD\_prior}.

%\section*{Acknowledgements}
%
%The author would like to thank Alexandra Schmidt and Milica Miocevic for their invaluable comments that helped improve the manuscript and Louis Dron for their help in extracting the data into an appropriate format for analysis and describing the study characteristics.

\bibliographystyle{authordate1}
{
\bibliography{ref01}
}

\newpage
\small{
	{
		\begin{table}
			\label{tab:1}
			\begin{tabular}{ L{4cm} c c C{3cm} c }
				& INTEREST & ZODIAC & PROCLAIM & Study 57\\
				\hline
				\hline
				Control group median overall survival (months) & 8 & 10 & 25 & 7.8\\
				\hline
				Stage III, (\%) & 38 & 15 & 100 & 17 \\
				\hline
				Stage IV, (\%)&53&85&0&83\\
				\hline
				Average age (years) &60.5&59&59&61\\
				\hline
				Adenocarcinoma histology (\%)&54&60&75&60\\
				\hline
				Two or more prior chemotherapy regimens (\%)&16&0&0&35\\
				\hline
				Radiotherapy sequence, dose (control arm)&None&None&60-66Gy, Concurrent&None
				
			\end{tabular}
			\caption{Summary of key trial characteristics for the four NSCLC trials}
		\end{table}
	}
}

%\begin{figure}[t]
%	\centering
%	\includegraphics[width=.75\textwidth]{example1.png}
%	\caption{Hypothetical example representing partially overlapping study populations.}
%	\label{fig:toy}
%\end{figure}


\end{document}